\documentclass[11pt, a4paper]{article}
\usepackage{amssymb}
\usepackage{eurosym}
\usepackage{amsfonts}
\usepackage{amsmath}
\usepackage{array}
\usepackage{calc}
\usepackage{array}
\usepackage{indentfirst}
\usepackage{graphicx}
\usepackage{hyperref}

\usepackage{authblk}
\author[1]{Remigiusz Durka \thanks{remigiusz.durka@uwr.edu.pl}}
\author[1]{Kamil Grela \thanks{kamilgrela97@gmail.com}}
\affil[1]{Institute for Theoretical Physics, University of Wroc\l{}aw, pl.\ M.\ Borna 9, 50-204 Wroc\l{}aw, Poland}
\title{On the number of possible resonant algebras}
\date{\today}

\begin{document}
	\maketitle
\begin{abstract}

We investigate the number of distinct resonant algebras depending on the generator content, which consists of the Lorentz generator, translation, and new additional Lorentz-like and translation-like generators. Such algebra enlargements originate directly from the Maxwell algebra and implementation of the S-expansion framework. Resonant algebras, being sub-class of the S-expanded algebras, should find use in the construction of gravity and supergravity models along some other applications. The undertaken task of establishing all the possible resonant algebras is closely related to the subject of finding commutative monoids (semigroups with the identity element) of a particular order, were we additionally enforce the parity condition.
 
\end{abstract}

\section{Introduction}
	
Over the past few years we have observed the growing number of works exploring the framework of the semigroup expansion (S-expansion) \cite{Izaurieta:2006zz, Diaz:2012zza, Salgado:2014qqa}. The resulting enlargements of the Lorentz/(A)dS/Poincar\'e algebras correspond to the so-called \textit{Maxwell algebras} containing the original Maxwell algebra introduced in the 70's \cite{Bacry:1970ye, Schrader:1972zd}, the Soroka-Soroka algebra \cite{Soroka:2004fj,Soroka:2006aj}, as well as their generalizations to $\mathfrak{B}_m$ and $\mathfrak{C}_m$ families \cite{Salgado:2014qqa, Concha:2016kdz}, along with extension to the new family $\mathfrak{D}_m$ \cite{Concha:2016hbt}, and a recipe for further families \cite{Durka:2016eun}. Ultimately, one can generalize it even further to the wider class of the so-called \textit{resonant algebras} \cite{Durka:2016eun, Durka:2019guk}. 

Mentioned enlargements deliver rich structure offering some non-trivial and interesting features. Their applications correspond to various aspects of the construction of the gravity and supergravity actions. Notable examples include \cite{Edelstein:2006se,deAzcarraga:2010sw,Penafiel:2018vpe,Fierro:2014lka,Kamimura:2011mq,deAzcarraga:2012zv, Concha:2014vka, Concha:2016zdb, Concha:2018zeb, Aviles:2018jzw,Inostroza:2014vua,Hoseinzadeh:2014bla,Salgado:2014jka,Durka:2011nf,Durka:2011gm,Durka:2011va,Durka:2012wd}. In particular, it is worthwhile to mention a recent contribution to the subject of $\mathfrak{bms}$ symmetry \cite{Concha:2018jxx, Concha:2018jjj,Salgado-Rebolledo:2019kft} and topological insulators \cite{Durka:2019guk, Palumbo:2016nku}. 

In the indicated constructions of the gauge theories, the particular features and outcomes depend on the chosen Lie algebra, which has a direct impact on the field content, structure constants, symmetries, existence of particular terms in the action and field equations. Therefore, determining the range of possible algebras for a given generator content is of great importance. We are going to tackle this problem by employing computer-assisted computations. After a brief introduction to the semigroup expansion, leading to the discussed algebra enlargements, we will present the outcomes and provide some classification. 
 
\section{Semigroup expansion}
	
In this section, we will briefly summarize the key elements of the semigroup expansion. First, let us recall the \text{İnönü}-Wigner contraction \cite{Inonu:1953sp} allowing us to obtain the Poincar\'{e} algebra as the limit of the Anti-de Sitter (AdS) algebra. This is established by the re-scaling of the translation generator $P_a\to \ell P_a$ with the constant $\Lambda=-\frac{3}{\ell^2}$. The commutator $[P_a,P_b]=J_{ab}$ then becomes $[\ell P_a,\ell P_b]=J_{ab}$ and $[P_a,P_b]=\frac{1}{\ell^2}J_{ab}$. Enforcing the limit corresponding to the vanishing of cosmological constant makes the right-hand side equal to zero
\begin{equation}
[P_a,P_b]\overset{\mathrm{\ell\to \infty}}{=}0\,.
\end{equation}
This way we can connect two different algebras. The semigroup expansion framework can be seen as a generalization, which assigns the semigroup elements to generators instead of a scalar parameter like $\ell$. This new "semigroup scaling" now concerns all the generators. Note, however, that various algebraic outcomes on the right-hand sides are not the consequence of a limit, but the generator redefinition after the evaluation of the original AdS commutators. The choice of the AdS and not dS is due to the possible supersymmetric applications.

The S-expanded algebra (see \cite{Izaurieta:2006zz, Diaz:2012zza, Salgado:2014qqa} and \cite{Edelstein:2006se}) is understood as the product of $S \times \mathfrak{g}$, where the new generators are given by:
\begin{align}
J_{ab,(i)}=s_{2i}\tilde{J}_{ab}\qquad \text{and}\qquad P_{a,(i)}=s_{2i+1}\tilde{P}_{a},\qquad \text{for}~i=\{0,1,2,...\}
\end{align}
with the original algebra $\mathfrak{g}=AdS$:
\small
\begin{align}
\left[ \tilde{J}_{ab},\tilde{J}_{cd}\right]   & =\eta_{bc}\tilde{J}_{ad}-\eta_{ac}\tilde{J}_{bd}+\eta
_{ad}\tilde{J}_{bc}-\eta_{bd}\tilde{J}_{ac}\,,\\
\left[\tilde{J}_{ab},\tilde{P}_{c}\right] & =\eta_{bc}\tilde{P}_{a}-\eta_{ac}\tilde{P}_{b}\,,\\
\left[\tilde{P}_{a},\tilde{P}_{b}\right] & =\tilde{J}_{ab}\,,
\end{align}
\normalsize
and semigroup elements $s_i \in S$. All the commutation relations between the generators of the enlarged algebra are obtained from the AdS algebra spanned by $\tilde{J}_{ab}$ and $\tilde{P}_a$ along the particular semigroup $S=\{s_0,s_1,s_2,...\}$ multiplication (Cayley) table
\begin{equation}
\begin{tabular}{c|cccc}
& $s_0$ & $s_1$ & $s_2$ & $.$ \\ \hline
$s_0$ & . & . & . & . \\
$s_1$ & . & . & . & . \\
$s_2$ & . & . & . & . \\
$.$   & . & . & . & .
\end{tabular}
\end{equation}
Elements in such table must admit commutative property $s_\alpha\cdot s_\beta=s_\beta\cdot s_\alpha$ and associativity condition $(s_\alpha\cdot s_\beta) \cdot s_\gamma=s_\alpha\cdot (s_\beta \cdot s_\gamma)$. The associativity directly translates to the Jacobi identity  $[[X,Y],Z]+[[Y,Z],X]+[[Z,X],Y]=0$ at the level of the generators. For a convenience in the further part of the article, we will name the generators using distinct letters: 
\begin{align}
J_{ab}= s_0\tilde{J}_{ab}\,,\quad P_{a}= s_1\tilde{P}_a\,,\quad Z_{ab}= s_2\tilde{J}_{ab}\,,\quad R_{a}= s_3\tilde{P}_a\,,\quad W_{ab}= s_4\tilde{J}_{ab}\,,\quad...
\end{align}

Physically useful algebras, from the point of view of the gauge actions, require the Lorentz generator obeying $[J, J]\sim J$ and $[J, P]\sim P$, so we do not break the definitions of the curvature form $R^{ab}(\omega)=d\omega^{ab}+\omega^{a~}_{~c} \wedge \omega^{cb}$ and torsion $T^{a}=d e^{a}+\omega^{a~}_{~c} \wedge e^{c}$ \cite{Concha:2016hbt,Durka:2016eun}. This should be extended further to all the other generators. Therefore, we start by introducing a semigroup element $s_0$ playing the role of the identity element  $s_0\cdot s_\alpha=s_\alpha\cdot s_0=s_\alpha$, which needs to be associated with the Lorentz generator $J_{ab}$ preserving all the generators $[J, X]\sim X$. Similarly, we associate $s_{1}$ element with the translation generator. In addition, we require the so-called resonant condition (see \cite{Izaurieta:2006zz, Diaz:2012zza, Salgado:2014qqa}), which mathematically is a parity requirement:
\begin{align}
s_{even}\cdot s_{even}=s_{even}\,,\qquad s_{even}\cdot s_{odd}=s_{odd}\,,\qquad s_{odd}\cdot s_{odd}=s_{even},
\end{align}
reflecting the AdS algebra structure
\begin{align}
[\tilde{J}_{..}, \tilde{J}_{..}]\sim \tilde{J}_{..}\qquad [\tilde{J}_{..}, \tilde{P}_{.}]\sim \tilde{P}_{.} \qquad [\tilde{P}_{.}, \tilde{P}_{.}]\sim \tilde{J}_{..} \, .\label{AdS}
\end{align}
To complete the picture, we also include the absorbing element $0_S$ separately, defined as $0_S \cdot s_\alpha=s_\alpha \cdot 0_S=0_S$. It is not related to any generator. On the contrary, by acting on any generator $T$ it gives $0_S\, T=0$, which is needed for example as the output of the Poincar\'e algebra. More on this subject can be found under the name of $0_S$-reduction \cite{Izaurieta:2006zz, Diaz:2012zza, Salgado:2014qqa}. While providing the algebra systematization, to make the semigroup order $n$ and the number of the generators on the same level, we will not include the separate row and column for $0_S$ in semigroup/monoid Cayley tables, as they do not contain anything else but $0_S$ elements.

To better understand the presented framework, let us look at one particular example of the original Maxwell algebra \cite{Bacry:1970ye, Schrader:1972zd} with the generators $J_{ab},P_{a},Z_{ab}$. Keeping in mind the specific form of structure constants, one can read all the commutators directly from the schematic "commutation" table
\begin{equation}
\begin{tabular}{c|lll}
	$[\,.\,,.\,]$& $J_{..}$ & $P_{.}$ & $Z_{..}$ \\ \hline
	$J_{..}$ & $J_{..}$ & $P_{.}$ & $Z_{..}$ \\
	$P_{.}$ & $P_{.}$ & $Z_{..}$ & $0$ \\
	$Z_{..}$ & $Z_{..}$ & $0$ & $0$
\end{tabular}
\end{equation}
This is a consequence of the corresponding semigroup \textit{multiplication} table
\begin{equation}
\begin{tabular}{c|ccc}
\framebox{$\mathfrak{B}_{4}$}& $s_0$ & $s_1$ & $s_2$ \\ \hline
$s_0$ & $s_0$ & $s_1$ & $s_2$ \\
$s_1$ & $s_1$ & $s_2$ & $0_S$ \\
$s_2$ & $s_2$ & $0_S$ & $0_S$
\end{tabular}
\end{equation}
applied to
\small
\begin{align}
\left[  J_{ab},J_{cd}\right]   &  = s_0 \cdot s_0(\eta_{bc}\tilde{J}_{ad}-\eta_{ac}\tilde{J}_{bd}+\eta
_{ad}\tilde{J}_{bc}-\eta_{bd}\tilde{J}_{ac})=\eta_{bc}J_{ad}-\eta_{ac}J_{bd}+\eta
_{ad}J_{bc}-\eta_{bd}J_{ac}\,,\nonumber\\
\left[  J_{ab},Z_{cd}\right]   &  = s_0 \cdot s_2(\eta_{bc}\tilde{J}_{ad}-\eta_{ac}\tilde{J}_{bd}+\eta
_{ad}\tilde{J}_{bc}-\eta_{bd}\tilde{J}_{ac})=\eta_{bc}Z_{ad}-\eta_{ac}Z_{bd}+\eta
_{ad}Z_{bc}-\eta_{bd}Z_{ac}\,,\nonumber\\
\left[  Z_{ab},Z_{cd}\right]   &  = s_2 \cdot s_2(\eta_{bc}\tilde{J}_{ad}-\eta_{ac}\tilde{J}_{bd}+\eta
_{ad}\tilde{J}_{bc}-\eta_{bd}\tilde{J}_{ac})=0\,,\nonumber\\
\left[  J_{ab}%
,P_{c}\right] & = s_0 \cdot s_1(\eta_{bc}\tilde{P}_{a}-\eta_{ac}\tilde{P}_{b})=\eta_{bc}P_{a}-\eta_{ac}P_{b}\,,\nonumber\\
\left[  Z_{ab}%
,P_{c}\right] & = s_2 \cdot s_1(\eta_{bc}\tilde{P}_{a}-\eta_{ac}\tilde{P}_{b})=0\,,\nonumber\\
\left[  P_{a},P_{b}\right]   & = s_1 \cdot s_1\tilde{J}_{ab}=Z_{ab}\,.
\end{align}
\normalsize
The internal product/invariant tensor $\left<\, .\, , .\, \right>$, necessary for the construction of actions, can be similarly read off, according to the same scheme with the particular constants $\sigma_\gamma$, depending on the outcome $s_\alpha \cdot s_\beta=s_\gamma$. For example, in 3D theory with S-expanded algebra with $s_2\cdot s_1=s_3$, we will have $\left<Z_{ab},P_c\right>=\sigma_3\epsilon_{abc}$, and so on. Naturally, when one considers $3D$ theory (like in \cite{Durka:2019guk}) it is also quite convenient to make a transition to the dual definitions of the generators $X_a=\frac{1}{2}\epsilon_{a~~}^{~bc}X_{bc}$ (and the corresponding fields), which simplifies and unifies the the commutation relations
\begin{align}
[J_a,J_b]=\epsilon_{ab~}^{~~~c}J_c\, ,\qquad [J_a,P_b]=\epsilon_{ab~}^{~~~c}P_c\, ,\qquad[P_a,P_b]=\epsilon_{ab~}^{~~~c}J_c\,.
\end{align}
More details about derivation, notation, and gauging these type of algebras can be found in the literature concerning the S-expansion \cite{Izaurieta:2006zz, Diaz:2012zza, Salgado:2014qqa} and \cite{Concha:2016kdz,Durka:2016eun, Durka:2019guk}.

\section{How many resonant algebras are there}

Some general considerations about finding all (unrestricted) semigroup expanded algebras were explored in the past \cite{Andrianopoli:2013ooa,Inostroza:2017ezc, Inostroza:2018uog}. Unfortunately, the overwhelming vastness of algebraic examples does not seem to translate into physically relevant results. We will argue that the key to assure the consistent grounds lies in demanding specific properties concerning the commutative semigroups. This includes the resonant condition and requirement of the identity element set to be $s_0$. Obtained in this way a class of \textit{resonant algebras} \cite{Durka:2016eun}, being sub-class of the S-expanded algebras, requires then a specific construction of the semigroups. To this end, we notice that the \textit{parity even} entries can have only $s_{2i}\cup\, 0_S$ elements related to $\{0, J, Z, W,...\}$, whereas \textit{parity odd} entries can be filled by odd $s_{2i+1}\cup\, 0_S$ related to $\{0, P, R, U, ...\}$. By brute-force we can generate all the possible \textit{multiplication} tables obeying given requirements with the exception of associativity. The associativity check has to be employed afterwards upon obtained candidates mathematically representing the “unital magma". The number of potential tables can be expressed by the combinatorial formula
\begin{align}
\left(\frac{2 k+5-(-1)^{k}}{4}\right)^{Floor\left[\frac{k^{2}}{4}\right]} \cdot \left(\frac{2 k+3+(-1)^{k}}{4}\right)^{Floor\left[\frac{(k-1)^{2}}{4}\right]}\,\label{candidates}
\end{align}
where $k$ represents the number of used generators. Starting point, $k=2$, corresponds to the $J,P$ setup. For $k=\{2,3,..., 8\}$ this gives us, respectively, 
$$2,18,729,331776,1073741824,64000000000000,37252902984619140625\,.$$
Associativity filtering reduces the final number of tables significantly, ultimately assuring the required commutative semigroup structure with identity and resonant condition. Based on them we generate all the resonant algebras. For the available generators $\{J_{ab},P_{a},Z_{ab},R_{a},W_{ab},U_{a}\}$, therefore up to three copies of the Lorentz/translation-like generators, we can summarize our explicit findings as
$$
\begin{tabular}{|c|c|c|c|c|c|c} \hline
& $\{J,P\}$ & $\{J,P,Z\}$ & $\{J,P,Z,R\}$ & $\{J,P,Z,R,W\}$ & $\{J,P,Z,R,W,U\}$  \\ \hline \hline
Possible algebras & 2 & 6 & 30 & 347  &  3786 \\ \hline
Algebras without 0 & 1 & 1 & 6 & 28  & 222 \\ \hline
\end{tabular}
$$
Naturally, a pair of $\{J_{ab}, P_{a}\}$ results in the Poincar\'{e} and AdS algebra. By enlarging algebra to a set of $\{J_{ab}, P_{a}, Z_{ab}\}$ we end up with the Maxwell algebra \cite{Bacry:1970ye, Schrader:1972zd} along with the Soroka-Soroka algebra \cite{Soroka:2004fj, Soroka:2006aj} and four more examples showed in \cite{Durka:2016eun} and \cite{Durka:2019guk}. The explicit form of algebras, up to the case of  $ \{J_{ab}, P_{a}, Z_{ab}, R_{a}\}$, will be presented in the next section, whereas rest will be attached to the publication in a separate file. Due to the plethora of cases, we suggest placing more emphasis on the fully non-abelian algebras, i.e. not having any zeros in the commutation outcomes, thus offering the most general content and the widest possible scope of the corresponding phenomena.

Before we go any further let us try to classify obtained resonant algebras. Depending on  $[P_a,P_b]$ outcome we can organize algebras according to the resulting sub-algebras:
$$
\begin{tabular}{|c|c|c|c|c|c|c|} \hline
& $\{J,P\}$ & $\{J,P,Z\}$ & $\{J,P,Z,R\}$ & $\{J,P,Z,R,W\}$ & $\{J,P,Z,R,W,U\}$  \\ \hline 
Possible algebras & 2 & 6 & 30 & 347  &  3786 \\ \hline\hline
Poincar\'e-like & 1 & 4 & 17 & 211 & 2062 \\ \hline
AdS-like & 1 & 0 & 3 & 0 & 38  \\ \hline
Maxwell-like & - & 2 & 10 & 68 &  843 \\ \hline
Extra-Maxwell-like & - & - & - & 68 &  843  \\ \hline
\end{tabular}
$$
Obviously Poincar\'e-like algebras are the ones possessing $[P_a,P_b]=0$, AdS-like $[P_a,P_b]=J_{ab}$, Maxwell-like $[P_a,P_b]=Z_{ab}$, and so on. 

The task of generating the resonant algebras is inextricably linked to the topic of finding the semigroups (and monoids) of a particular order $n$. Unfortunately, little is known about the inclusion of the parity condition. Observed numbers of particular resonant algebras seem not to follow any general sequence, just as in the case of the semigroups and monoids. In literature \cite{Andrianopoli:2013ooa, Forsythe:1955, Motzkin:1956, Plemmons:1967, Distler:2009, Distler:2012} and through \href{https://oeis.org/}{oeis.org} we find:
\begin{itemize}
\item A001423 \quad Number of semigroups of order $n$ for $n=1,2,3,4,5,6,7,8,9,10$:\\
$1,4,18,126,1160,15973,836021,1843120128,52989400714478,12418001077381302684 $
\item A001426 \quad Number of commutative semigroups of order $n$:\\
$1,3,12,58,325,2143,17291,221805,11545843,3518930337$
\item A058129 \quad Number of monoids (semigroups with identity) of order $n$:\\
$1,2,7,35,228,2237,31559,1668997,...$
\item A058131 \quad Number of commutative monoids of order $n$:\\
$1,2,5,19,78,421,2637,...$
\end{itemize}
One might wonder why not start with a given (commutative) semigroups available up to order $n=10$, and then simply filter them for the identity and parity. This path naturally offers much higher efficiency paid by someone's previous work. Unfortunately, the available tables are given up to isomorphism. For example, with the identity set as the first element and without the resonant condition, we can find 9 commutative monoids of the order $n=3$:
\begin{align*}
\footnotesize
\begin{tabular}{c|ccc}
$M_1$ & $s_0$ & $s_1$ & $s_2$ \\ \hline
$s_0$ & $s_0$ & $s_1$ & $s_2$ \\
$s_1$ & $s_1$ & $s_2$ & $s_0$ \\
$s_2$ & $s_2$ & $s_0$ & $s_1$
\end{tabular}
~~
\begin{tabular}{c|ccc}
$M_2$ & $s_0$ & $s_1$ & $s_2$ \\ \hline
$s_0$ & $s_0$ & $s_1$ & $s_2$ \\
$s_1$ & $s_1$ & $s_0$ & $s_2$ \\
$s_2$ & $s_2$ & $s_2$ & $s_2$
\end{tabular}
~~
\begin{tabular}{c|ccc}
$M_3$ & $s_0$ & $s_1$ & $s_2$ \\ \hline
$s_0$ & $s_0$ & $s_1$ & $s_2$ \\
$s_1$ & $s_1$ & $s_1$ & $s_1$ \\
$s_2$ & $s_2$ & $s_1$ & $s_2$
\end{tabular}
~~
\begin{tabular}{c|ccc}
$M_4$ & $s_0$ & $s_1$ & $s_2$ \\ \hline
$s_0$ & $s_0$ & $s_1$ & $s_2$ \\
$s_1$ & $s_1$ & $s_1$ & $s_1$ \\
$s_2$ & $s_2$ & $s_1$ & $s_1$
\end{tabular}
~~
\begin{tabular}{c|ccc}
$M_5$ & $s_0$ & $s_1$ & $s_2$ \\ \hline
$s_0$ & $s_0$ & $s_1$ & $s_2$ \\
$s_1$ & $s_1$ & $s_1$ & $s_2$ \\
$s_2$ & $s_2$ & $s_2$ & $s_1$
\end{tabular}
\\[5pt]\footnotesize
\begin{tabular}{c|ccc}
$M_6$ & $s_0$ & $s_1$ & $s_2$ \\ \hline
$s_0$ & $s_0$ & $s_1$ & $s_2$ \\
$s_1$ & $s_1$ & $s_1$ & $s_1$ \\
$s_2$ & $s_2$ & $s_1$ & $s_0$
\end{tabular}
~~
\begin{tabular}{c|ccc}
$M_7$ & $s_0$ & $s_1$ & $s_2$ \\ \hline
$s_0$ & $s_0$ & $s_1$ & $s_2$ \\
$s_1$ & $s_1$ & $s_1$ & $s_2$ \\
$s_2$ & $s_2$ & $s_2$ & $s_2$
\end{tabular}
~~
\begin{tabular}{c|ccc}
$M_8$ & $s_0$ & $s_1$ & $s_2$ \\ \hline
$s_0$ & $s_0$ & $s_1$ & $s_2$ \\
$s_1$ & $s_1$ & $s_2$ & $s_2$ \\
$s_2$ & $s_2$ & $s_2$ & $s_2$
\end{tabular}
~~
\begin{tabular}{c|ccc}
$M_9$ & $s_0$ & $s_1$ & $s_2$ \\ \hline
$s_0$ & $s_0$ & $s_1$ & $s_2$ \\
$s_1$ & $s_1$ & $s_2$ & $s_1$ \\
$s_2$ & $s_2$ & $s_1$ & $s_2$
\end{tabular}
\end{align*}
\normalsize
The four lower examples will be discarded, because after redefining the elements $s_1\leftrightarrow s_2$ and changing the order of rows and columns we reproduce exactly four examples lying above them. We could even miss the Soroka-Soroka $\mathfrak{C}_4$ algebra, represented by the last $M_9$ table (incidentally, the only one obeying the resonant condition!) as it might not be directly visible in an isomorphic output. The presence of the absorbing element introduces yet another complication. Including $0_S$ means searching among higher order monoids. Notice that $M_8$ table gives the Poincar\'{e} algebra with $s_2$ not related to the $Z_{ab}$ generator, but playing the role of $0_S$ element.

The setup presented at the beginning of this section is much more convenient. The physical properties associated with $s_0$ and $s_1$, along with the resonant/parity condition, remove the isomorphic ambiguities, because similar re-definitions as mentioned above are just impossible. The lack of physical restrictions concerning $Z_{ab}$ and $W_{ab}$ results in interchangeability that might lead to some classification problems. For now, we will assume that the role of all generators is somehow unique. 

\section{Panorama of algebras}

Below we present an overview of the resonant algebras obtained by subsequent inclusion of further generators.
\subsection{J algebra}
We can consider the Lorentz algebra as the starting point obeying all the requirements:
\small
\begin{equation}
\begin{tabular}{c|cc}
Lorentz & J \\ \hline
J & J
\end{tabular}
\end{equation}
\normalsize
\subsection{J and P algebras}
With the translation generator we see only two algebra possibilities, i.e. Poincar\'e and AdS:
\small
\begin{equation}
\begin{tabular}{c|cc}
Poincar\'{e} & J & P \\ \hline
J & J & P  \\
P & P & 0 
\end{tabular}
\qquad
\begin{tabular}{c|cc}
AdS & J & P \\ \hline
J & J & P  \\
P & P & J 
\end{tabular}
\end{equation}
\normalsize
\subsection{J and P and Z algebras}
Including another generator, $Z_{ab}$, brings much richer structure:
\begin{itemize}
	\item 2 Maxwell-like algebras (i.e. containing $[P,P]\sim Z$): of type $\mathfrak{B}_{4}$ (original
	Maxwell algebra introduced in the 70's) and type $\mathfrak{C}_{4}\equiv
	AdS\oplus Lorentz$ (introduced by Soroka-Soroka, which was shown to represent under a change of basis the direct sum of two algebras)
	\small
	\begin{equation}
	\begin{tabular}{c|ccc}
	\framebox{$\mathfrak{B}_{4}$}& J & P & Z \\ \hline
	J & J & P & Z \\
	P & P & Z & 0 \\
	Z & Z & 0 & 0
	\end{tabular}
	\qquad
	\begin{tabular}{c|ccc}
	\framebox{$\mathfrak{C}_{4}$}& J & P & Z \\ \hline
	J & J & P & Z \\
	P & P & Z & P \\
	Z & Z & P & Z
	\end{tabular}
	\end{equation}
	\normalsize
	\item none of AdS-like algebras, which would contain $[P,P]\sim J$
	\item 4 Poincar\'{e}-like algebras (i.e. having $[P,P]\sim 0$), which we denote as type: $B_{4}$, $%
	\tilde{B}_{4}$, $\tilde{C}_{4}$, and $C_{4}\equiv Poincare\oplus Lorentz$
	\small
	\begin{equation}
	\begin{tabular}{c|ccc}
	\framebox{$B_{4}$} & J & P & Z \\ \hline
	J & J & P & Z \\
	P & P & 0 & 0 \\
	Z & Z & 0 & 0
	\end{tabular}
	\quad
	\begin{tabular}{c|ccc}
	\framebox{$\tilde{B}_{4}$}& J & P & Z \\ \hline
	J & J & P & Z \\
	P & P & 0 & P \\
	Z & Z & P & 0
	\end{tabular}
	\quad
	\begin{tabular}{c|ccc}
	\framebox{$\tilde{C}_{4}$}& J & P & Z \\ \hline
	J & J & P & Z \\
	P & P & 0 & 0 \\
	Z & Z & 0 & Z
	\end{tabular}
	\quad
	\begin{tabular}{c|ccc}
	\framebox{$C_{4}$} & J & P & Z \\ \hline
	J & J & P & Z \\
	P & P & 0 & P \\
	Z & Z & P & Z
	\end{tabular}
	\end{equation}
\end{itemize}
\normalsize
Altogether for the set of three generators $\{J,P,Z\}$ we have thus 6 different algebras. As was recently shown in the analysis of topological insulators \cite{Durka:2019guk}, they correspond to six different Lagrangians and therefore six different configurations of the field equations.

\subsection{J and P and Z and R algebras}
The last example presented here explicitly consists of 30 cases:
\begin{itemize}
	\item 10 Maxwell-like
	\small
	\begin{align}
	\begin{tabular}{c|cccc}
	\framebox{$\mathfrak{B}_{5}$} & J & P & Z & R \\ \hline
	J & J & P & Z & R \\ 
	P & P & Z & R & 0 \\ 
	Z & Z & R & 0 & 0 \\ 
	R & R & 0 & 0 & 0%
	\end{tabular}%
	\qquad 
	\begin{tabular}{c|cccc}
	\framebox{$\mathfrak{C}_{4}$} & J & P & Z & R \\ \hline
	J & J & P & Z & R \\ 
	P & P & Z & R & J \\ 
	Z & Z & R & J & P \\ 
	R & R & J & P & Z%
	\end{tabular}%
	\qquad 
	\begin{tabular}{c|cccc}
	\framebox{$\mathfrak{D}_{4}$} & J & P & Z & R \\ \hline
	J & J & P & Z & R \\ 
	P & P & Z & R & Z \\ 
	Z & Z & R & Z & R \\ 
	R & R & Z & R & Z%
	\end{tabular}%
	\nonumber\\
	\begin{tabular}{c|cccc}
	\qquad\qquad & J & P & Z & R \\ \hline
	J & J & P & Z & R \\ 
	P & P & Z & P & 0 \\ 
	Z & Z & P & Z & 0 \\ 
	R & R & 0 & 0 & 0%
	\end{tabular}%
	\qquad 
	\begin{tabular}{c|cccc}
	\qquad\qquad & J & P & Z & R \\ \hline
	J & J & P & Z & R \\ 
	P & P & Z & P & Z \\ 
	Z & Z & P & Z & P \\ 
	R & R & Z & P & J%
	\end{tabular}%
	\qquad 
	\begin{tabular}{c|cccc}
	\qquad\qquad & J & P & Z & R \\ \hline
	J & J & P & Z & R \\ 
	P & P & Z & P & Z \\ 
	Z & Z & P & Z & P \\ 
	R & R & Z & P & Z%
	\end{tabular}
	\nonumber
	\end{align}
	\begin{align}
	\begin{tabular}{c|cccc}
	\qquad\qquad  & J & P & Z & R \\ \hline
	J & J & P & Z & R \\ 
	P & P & Z & 0 & 0 \\ 
	Z & Z & 0 & 0 & 0 \\ 
	R & R & 0 & 0 & 0%
	\end{tabular}%
	\qquad 
	\begin{tabular}{c|cccc}
	\qquad\qquad  & J & P & Z & R \\ \hline
	J & J & P & Z & R \\ 
	P & P & Z & 0 & 0 \\ 
	Z & Z & 0 & 0 & 0 \\ 
	R & R & 0 & 0 & Z%
	\end{tabular}%
	\nonumber\\
	\begin{tabular}{c|cccc}
	\qquad\qquad  & J & P & Z & R \\ \hline
	J & J & P & Z & R \\ 
	P & P & Z & 0 & Z \\ 
	Z & Z & 0 & 0 & 0 \\ 
	R & R & Z & 0 & 0%
	\end{tabular}%
	\qquad 
	\begin{tabular}{c|cccc}
	\qquad\qquad  & J & P & Z & R \\ \hline
	J & J & P & Z & R \\ 
	P & P & Z & 0 & Z \\ 
	Z & Z & 0 & 0 & 0 \\ 
	R & R & Z & 0 & Z%
	\end{tabular}%
	\end{align}
	\normalsize
	\item 3 AdS-like
	\small
	\begin{align}
	\begin{tabular}{c|cccc}
	\framebox{$\mathcal{B}_{5}$} & J & P & Z & R \\ \hline
	J & J & P & Z & R\\
	P & P & J & R & Z\\
	Z & Z & R & 0 & 0\\
	R & R & Z & 0 & 0\\
	\end{tabular}
	\qquad
	\begin{tabular}{c|cccc}
	\framebox{$\mathcal{C}_{5}$}  & J & P & Z & R \\ \hline
	J & J & P & Z & R\\
	P & P & J & R & Z\\
	Z & Z & R & J & P\\
	R & R & Z & P & J\\
	\end{tabular}
	\qquad
	\begin{tabular}{c|cccc}
	\framebox{$\mathcal{D}_{5}$} & J & P & Z & R \\ \hline
	J & J & P & Z & R\\
	P & P & J & R & Z\\
	Z & Z & R & Z & R\\
	R & R & Z & R & Z\\
	\end{tabular}
	\end{align}
	\normalsize
	\item 17 Poincar\'{e}-like
	\small 
	\begin{align}
	\begin{tabular}{c|cccc}
	\qquad\qquad & J & P & Z & R \\ \hline
	J & J & P & Z & R \\ 
	P & P & 0 & 0 & 0 \\ 
	Z & Z & 0 & 0 & 0 \\ 
	R & R & 0 & 0 & 0%
	\end{tabular}%
	\qquad 
	\begin{tabular}{c|cccc}
	\qquad\qquad & J & P & Z & R \\ \hline
	J & J & P & Z & R \\ 
	P & P & 0 & 0 & 0 \\ 
	Z & Z & 0 & Z & 0 \\ 
	R & R & 0 & 0 & 0%
	\end{tabular}%
	\qquad 
	\begin{tabular}{c|cccc}
	\qquad\qquad & J & P & Z & R \\ \hline
	J & J & P & Z & R \\ 
	P & P & 0 & 0 & 0 \\ 
	Z & Z & 0 & 0 & 0 \\ 
	R & R & 0 & 0 & Z%
	\end{tabular}
	\nonumber\\
	\begin{tabular}{c|cccc}
	\qquad\qquad & J & P & Z & R \\ \hline
	J & J & P & Z & R \\ 
	P & P & 0 & 0 & Z \\ 
	Z & Z & 0 & 0 & 0 \\ 
	R & R & Z & 0 & 0%
	\end{tabular}%
	\qquad 
	\begin{tabular}{c|cccc}
	\qquad\qquad & J & P & Z & R \\ \hline
	J & J & P & Z & R \\ 
	P & P & 0 & 0 & Z \\ 
	Z & Z & 0 & 0 & 0 \\ 
	R & R & Z & 0 & Z%
	\end{tabular}
	\qquad 
	\begin{tabular}{c|cccc}
	\qquad\qquad & J & P & Z & R \\ \hline
	J & J & P & Z & R \\ 
	P & P & 0 & R & 0 \\ 
	Z & Z & R & 0 & 0 \\ 
	R & R & 0 & 0 & 0%
	\end{tabular}%
	\nonumber\\
	\begin{tabular}{c|cccc}
	\qquad\qquad & J & P & Z & R \\ \hline
	J & J & P & Z & R \\ 
	P & P & 0 & 0 & 0 \\ 
	Z & Z & 0 & 0 & P \\ 
	R & R & 0 & P & 0%
	\end{tabular}%
	\qquad 
	\begin{tabular}{c|cccc}
	\qquad\qquad & J & P & Z & R \\ \hline
	J & J & P & Z & R \\ 
	P & P & 0 & P & 0 \\ 
	Z & Z & P & Z & 0 \\ 
	R & R & 0 & 0 & 0%
	\end{tabular}%
	\qquad 
	\begin{tabular}{c|cccc}
	\qquad\qquad & J & P & Z & R \\ \hline
	J & J & P & Z & R \\ 
	P & P & 0 & 0 & 0 \\ 
	Z & Z & 0 & Z & R \\ 
	R & R & 0 & R & 0%
	\end{tabular}
	\nonumber\\
	\begin{tabular}{c|cccc}
	\qquad\qquad & J & P & Z & R \\ \hline
	J & J & P & Z & R \\ 
	P & P & 0 & 0 & Z \\ 
	Z & Z & 0 & 0 & P \\ 
	R & R & Z & P & J%
	\end{tabular}%
	\qquad 
	\begin{tabular}{c|cccc}
	\qquad\qquad & J & P & Z & R \\ \hline
	J & J & P & Z & R \\ 
	P & P & 0 & 0 & 0 \\ 
	Z & Z & 0 & 0 & P \\ 
	R & R & 0 & P & Z%
	\end{tabular}%
	\qquad 
	\begin{tabular}{c|cccc}
	\qquad\qquad & J & P & Z & R \\ \hline
	J & J & P & Z & R \\ 
	P & P & 0 & 0 & 0 \\ 
	Z & Z & 0 & Z & R \\ 
	R & R & 0 & R & Z%
	\end{tabular}%
	\nonumber\\
	\begin{tabular}{c|cccc}
	\qquad\qquad & J & P & Z & R \\ \hline
	J & J & P & Z & R \\ 
	P & P & 0 & P & 0 \\ 
	Z & Z & P & J & R \\ 
	R & R & 0 & R & 0%
	\end{tabular}%
	\qquad 
	\begin{tabular}{c|cccc}
	\qquad\qquad & J & P & Z & R \\ \hline
	J & J & P & Z & R \\ 
	P & P & 0 & R & 0 \\ 
	Z & Z & R & J & P \\ 
	R & R & 0 & P & 0%
	\end{tabular}%
	\qquad 
	\begin{tabular}{c|cccc}
	\qquad\qquad & J & P & Z & R \\ \hline
	J & J & P & Z & R \\ 
	P & P & 0 & P & 0 \\ 
	Z & Z & P & Z & R \\ 
	R & R & 0 & R & 0%
	\end{tabular}%
	\nonumber\\~~\nonumber\\
	\begin{tabular}{c|cccc}
	\qquad\qquad & J & P & Z & R \\ \hline
	J & J & P & Z & R \\ 
	P & P & 0 & P & 0 \\ 
	Z & Z & P & Z & P \\ 
	R & R & 0 & P & 0%
	\end{tabular}%
	\qquad 
	\begin{tabular}{c|cccc}
	\qquad\qquad & J & P & Z & R \\ \hline
	J & J & P & Z & R \\ 
	P & P & 0 & R & 0 \\ 
	Z & Z & R & Z & R \\ 
	R & R & 0 & R & 0%
	\end{tabular}
	\qquad \qquad \qquad 
	\end{align}
	\normalsize
\end{itemize}

\subsection{Algebras with more generators}

We will not present explicitly all the 347 $\{J, P, Z, R, W\}$ algebras, nor 3876 examples of $\{J, P, Z, R, W, U\}$ but they will be available through a separately supplemented file. 

Values given in \eqref{candidates} show that incorporating further generators (i.e. beyond the last given pair of $\{W_{ab}, U_a\}$) becomes quite a time-consuming exercise, requiring in the next steps the associativity analysis of $6,4\times 10^{13}$ and $3,725 \times 10^{19}$ candidates. The last result of 3876 resonant algebras was established after $31$ hours of medium-class PC computations concerning "only" $10^9$ candidates.

\section{Other observations}

Besides the systematization over the appearing sub-algebras, one could focus on other aspects. More emphasis should be placed on the non-abelian algebras (without zeros in the commutator's outcome). They are particularly interesting because they give the most general actions. Intriguingly, we notice quite a small number of AdS-like algebras. It seems that we are unable to construct such an algebraic type for the odd number of generators. Generally, one would expect more in the future on the meaning of all additional bosonic fields.

Another issue concerns matching some numbers concerning the Maxwell-like and algebras of a further type.  Generators $Z_{ab}$ and $W_{ab}$ seem to be interchangeable from a certain point. This will be always the case, until we deliver some interpretation that will enforce special requirements for the $Z_{ab}$ generator similar to what we did for Lorentz and translation. 

Several resonant tables happen to form the \textit{cyclic group} $\mathbb{Z}_k$. There is one such table corresponding to $\{J, P\}$ content, one for $\{J, P, Z, R\}$, and then six for $\{J, P, Z, R, W, U\}$. Tables forming \textit{groups} (monoids with the inverses) require adding one more table with $\{J, P, Z, R\}$ to the list above. Analysis of the gauged actions show that the generators being related to each other by semigroup inverses do not translate into any physical implications.

The extended set of possible algebras calls for further analysis corresponding to the gauge models (Chern-Simons in odd dimensions, Born-Infeld in even dimensions \cite{Concha:2016kdz, Concha:2014vka}, or BF $\to$ BFCG construction \cite{Durka:2011nf, Durka:2012wd}). This also includes the supergravity models, where the supersymmetric extensions should be now explored for all the resonant algebras.

Finally, the setup with the two sets of Lorentz- and translation-like generators: $\{J, P, Z, R\}$ might be particularly interesting from the perspective of the bi-metric theories \cite{Hoseinzadeh:2017lkh}. The spin connection $\omega^{ab}$ and vielbein $e^a$ (associated with $J_{ab}$ and $P_a$) would correspond to the background metric, whereas the pair of fields $k^{ab}$ and $h^a$ (associated with the $Z_{ab}$ and $R_{a}$) could be related to the other metric field. Closing these algebras in thirty different ways means many non-trivial interaction (mixed) terms, elsewhere being constructed by hand. With more generators in play, this opens doors to the multi-metric formulation.

\section{Summary}

Within this paper, we have attempted to provide the complete overview of the \textit{resonant algebras} \cite{Durka:2019guk}, of which only a limited portion has been analyzed and incorporated in various applications. Provided scope and classification helps us better understand all the relations between algebras, which in the future might turn essential in realizing various formal and physical goals. 

\section{Acknowledgment}

Authors would like to thank B. Kosza\l ka for the past \href{https://resonantalgebras.wordpress.com/}{contribution} in generating the resonant algebras. Work was supported by the Institute Grant for Young Researchers 0420/2716/18.

\end{document}